\documentclass{article}
\usepackage{cite}
\usepackage{graphicx}
\usepackage{dcolumn}
\usepackage{fancyhdr}
\usepackage{appendix}

\input{tcilatex}
\begin{document}

\date{}
\title{Quasi-Hermitian one-dimensional lattice}
\author{Francisco M. Fern\'{a}ndez\thanks{%
fernande@quimica.unlp.edu.ar} \\
INIFTA, DQT, Sucursal 4, C. C. 16, \\
1900 La Plata, Argentina}
\maketitle

\begin{abstract}
We show that a non-Hermitian operator with a tridiagonal matrix
representation in a finite-dimensional vector space is similar to a
Hermitian operator. The required condition is sufficient and simple examples
show that it is not necessary. We derive quite general features of the
eigenvalues and eigenvectors for a somewhat particular case.
\end{abstract}

\section{Introduction}

\label{sec:intro}

Non-Hermitian quantum mechanics has become quite popular in recent years\cite
{BBJ03,B07} because of its intrinsic mathematical interest and as a suitable
tool for the interpretation of some physical phenomena. In particular,
exactly solvable models given in terms of tridiagonal matrices have proved
useful for deriving and illustrating some properties of non-Hermitian systems%
\cite{Z07a,Z07b,Z07c,Z08a,Z08b,Z09,Z10,Z11}. Some non-Hermitian operators
exhibit generalized Hermiticity\cite{P43} or quasi-Hermiticity\cite{SGH92}
that provides the condition for a linear operator to be similar to a
self-adjoint one\cite{W69}.

The purpose of this paper is a discussion of a non-Hermitian operator with a
finite tridiagonal matrix representation, similar to those discussed earlier%
\cite{Z07a,Z07b,Z07c,Z08a,Z08b,Z09,Z10,Z11}, though somewhat more general.
In section~\ref{sec:lattice} we derive the main result, in section~\ref
{sec:examples} we discuss some illustrative examples and in section~\ref
{sec:conclusions} we summarize the main results of this paper and draw
conclusions.

\section{Quasi-Hermitian one-dimensional lattice}

\label{sec:lattice}

Throughout this paper we consider the quantum-mechanical model given by the
Hamiltonian operator 
\begin{equation}
H=\sum_{j=1}^{N-1}\left( \alpha _{j}\left| j\right\rangle \left\langle
j+1\right| +\beta _{j}\left| j+1\right\rangle \left\langle j\right| \right)
+\sum_{j=1}^{N}\omega _{j}\left| j\right\rangle \left\langle j\right| ,
\label{eq:H}
\end{equation}
where $\left\{ \left| j\right\rangle ,\;j=1,2,\ldots ,N\right\} $ is an
orthonormal basis set. This operator is Hermitian if $\omega _{j}^{*}=\omega
_{j}$ and $\alpha _{j}=\beta _{j}^{*}$.

Any eigenvector $\left| \psi \right\rangle $ of $H$ is a linear combination
of the basis vectors 
\begin{equation}
\left| \psi \right\rangle =\sum_{j=1}^{N}c_{j}\left| j\right\rangle ,
\label{eq:eigenvector_psi}
\end{equation}
and it follows from the eigenvalue equation $H\left| \psi \right\rangle
=E\left| \psi \right\rangle $ that the coefficients $c_{j}$ satisfy the
three-term recurrence relation 
\begin{eqnarray}
\alpha _{j}c_{j+1}+\left( \omega _{j}-E\right) c_{j}+\beta _{j-1}c_{j-1}
&=&0,\;j=1,2,\ldots ,N,  \nonumber \\
c_{0} &=&c_{N+1}=0.  \label{eq:secular}
\end{eqnarray}

In what follows we resort to a mathematical argument used earlier by Child
et al\cite{CDW00} and Amore and Fern\'{a}ndez\cite{AF20} for the truncation
of some particular three-term recurrence relations. If we substitute $%
c_{j}=Q_{j}d_{j}$ into equation (\ref{eq:secular}) and divide by $Q_{j}$ we
obtain a secular equation for the new coefficients $d_{j}$%
\begin{equation}
\alpha _{j}\frac{Q_{j+1}}{Q_{j}}\,d_{j+1}+\left( \omega _{j}-E\right)
d_{j}+\beta _{j-1}\frac{Q_{j-1}}{Q_{j}}\,d_{j-1}=0.  \label{eq:secular_Q}
\end{equation}
The tridiagonal matrix that gives rise to this secular equation is symmetric
if 
\begin{equation}
Q_{j+1}^{2}=\frac{\beta _{j}}{\alpha _{j}}\,Q_{j}^{2}.
\label{eq:Q_j_relationship}
\end{equation}
Therefore, we conclude that if $\alpha _{j}$, $\beta _{j}$, and $\omega _{j}$
are real and $\alpha _{j}\beta _{j}>0$, then the tridiagonal matrix (\ref
{eq:secular_Q}) is symmetric and, consequently, all its eigenvalues are
real. In fact, if we choose $Q_{j+1}/Q_{j}=\sqrt{\beta _{j}/\alpha _{j}}$
then the secular equation (\ref{eq:secular_Q}) becomes 
\begin{equation}
\sqrt{\alpha _{j}\beta _{j}}\,d_{j+1}+\left( \omega _{j}-E\right) d_{j}+%
\sqrt{\alpha _{j-1}\beta _{j-1}}\,d_{j-1}=0.  \label{eq:secular_symmetric}
\end{equation}
Note that the boundary conditions have not changed because $%
c_{0}=c_{N+1}=0\Rightarrow d_{0}=d_{N+1}=0$.

If $\mathbf{c}$ is a column vector with elements $c_{j}$ and $\mathbf{H}$
the tridiagonal square matrix with elements $H_{ij}=\left\langle i\right|
H\left| j\right\rangle $ then $\mathbf{Hc}=E\mathbf{c}$ is the secular
equation (\ref{eq:secular}) in matrix form. The column vector $\mathbf{d}$,
with elements $d_{j}$, is related to $\mathbf{c}$ by means of the diagonal
invertible matrix $\mathbf{Q}$ with elements $Q_{ij}=Q_{j}\delta _{ij}$.
Therefore, we have 
\begin{equation}
\mathbf{Hc}=E\mathbf{c}\Rightarrow \mathbf{HQd}=E\mathbf{Qd\Rightarrow 
\mathbf{Q}^{-1}HQd}=E\mathbf{d},
\end{equation}
where the transformed matrix $\tilde{\mathbf{H}}=\mathbf{\mathbf{Q}^{-1}HQ}$
is symmetric. It follows from $\tilde{\mathbf{H}}^{t}=\tilde{\mathbf{H}}$,
where $t$ stands for transpose, that $\mathbf{HQ}^{2}=\mathbf{Q}^{2}\mathbf{H%
}^{t}$, where $\mathbf{Q}^{2}$ is symmetric, invertible and positive
definite. Clearly, the matrix $\mathbf{H}$ discussed here is an example of
the operator $T$ in the theorem proved by Williams\cite{W69}. The matrix
elements of $\mathbf{Q}^{2}$ are given by 
\begin{equation}
Q_{j}^{2}=Q_{1}^{2}\prod_{k=1}^{j-1}\frac{\beta _{j-k}}{\alpha _{j-k}},
\label{eq:Q_j^2}
\end{equation}
where $Q_{1}$ is any real nonzero number.

If we choose the metric $\mathbf{Q}^{-2}$\cite{Z10} (and references therein)
then the resulting eigenvectors $\mathbf{c}_{k}=\mathbf{Qd}_{k}$ can be
orthonormalized according to $\mathbf{c}_{j}^{t}\cdot \mathbf{Q}^{-2}\cdot 
\mathbf{c}_{k}=\mathbf{d}_{j}^{t}\mathbf{d}_{k}=\delta _{ij}$. If $\mathbf{D}%
=\left( \mathbf{d}_{1}\,\mathbf{d}_{2}\,\ldots \,\mathbf{d}_{N}\right) $ is
the $N\times N$ unitary matrix with the column vector $\mathbf{d}_{j}$ as
its $j$-th column ($\mathbf{D}^{\dagger }\mathbf{D}=\mathbf{I}$, where $%
\mathbf{I}$ is the $N\times N$ identity matrix), then $\mathbf{D}^{\dagger }%
\tilde{\mathbf{H}}\mathbf{D}=\mathbf{E}$, where $E_{ij}=E_{i}\delta _{ij}$.
Therefore, $\mathbf{S}^{-1}\mathbf{HS=E}$, where $\mathbf{S}=\mathbf{QD}$.

By means of the invertible Hermitian operator 
\begin{equation}
Q=\sum_{j=1}^{N}Q_{j}\left| j\right\rangle \left\langle j\right| ,
\label{eq:Q_op}
\end{equation}
we can define the Hermitian operator $\tilde{H}=Q^{-1}HQ$ that satisfies $%
HQ^{2}=Q^{2}H^{\dagger }$, where $\dagger $ stands for Hermitian conjugate.
Clearly, $H$ satisfies the hypothesis of Williams' theorem\cite{W69}.

We can easily derive a general result about the eigenvalues and eigenvectors
of the Hamiltonian operator (\ref{eq:H}) when $\omega _{j}=\omega $ for all $%
j$. In this case we can rewrite the secular equation (\ref{eq:secular}) as 
\begin{equation}
\alpha _{j}c_{j+1}-\epsilon c_{j}+\beta _{j-1}c_{j-1}=0,
\end{equation}
where $\epsilon =E-\omega $. If we substitute $c_{j}=(-1)^{j}\tilde{c}_{j}$
then this equation becomes 
\begin{equation}
\alpha _{j}\tilde{c}_{j+1}+\epsilon \tilde{c}_{j}+\beta _{j-1}\tilde{c}%
_{j-1}=0.
\end{equation}
We conclude that if $\mathbf{c}_{k}$ is an eigenvector of $\mathbf{H}$ with
eigenvalue $\epsilon _{k}$ then $\tilde{\mathbf{c}}_{k}$ is an eigenvector
with eigenvalue $-\epsilon _{k}$ and the eigenvalues $E_{k}$ of the
Hamiltonian (\ref{eq:H}) are symmetrically distributed about $\omega $.
Besides, if $N$ is odd then there is always an eigenvalue $E_{k}=\omega $
(or $\epsilon _{k}=0$).

\section{Exactly solvable examples}

\label{sec:examples}

The simplest example is 
\begin{equation}
\mathbf{H}=\left( 
\begin{array}{ll}
\omega _{1} & \alpha \\ 
\beta & \omega _{2}
\end{array}
\right) ,  \label{eq:H_2x2}
\end{equation}
with eigenvalues 
\begin{equation}
E_{\pm }=\frac{\omega _{1}+\omega _{2}\pm \sqrt{\left( \omega _{1}-\omega
_{2}\right) ^{2}+4\alpha \beta }}{2}.  \label{eq:E_2x2}
\end{equation}
We appreciate that these eigenvalues are real for all $\alpha \beta >-\left(
\omega _{1}-\omega _{2}\right) ^{2}/4$ which shows that the condition $%
\alpha \beta >0$ proposed in section~\ref{sec:lattice} is sufficient but not
necessary. When $\omega _{1}=\omega _{2}=\omega $ the eigenvalues $E_{\pm
}=\omega \pm \sqrt{\alpha \beta }$ exhibit the symmetric distribution about $%
\omega $ derived in section~\ref{sec:lattice} and are real when $\alpha
\beta >0$.

The second example is given by $\omega _{j}=\omega $, $\alpha _{j}=\alpha $, 
$\beta _{j}=\beta $ for all $j$. In this case, equation (\ref
{eq:secular_symmetric}) can be rewritten as 
\begin{equation}
d_{j-1}-x\,d_{j}+d_{j+1}=0,\;x=\frac{E-\omega }{\sqrt{\alpha \beta }}.
\label{eq:secular_Huckel}
\end{equation}
This simple H\"{u}ckel-like equation\cite{P68} can be easily solved and the
result is 
\begin{eqnarray}
E_{k} &=&\omega +2\sqrt{\alpha \beta }\cos \left( \frac{k\pi }{N+1}\right)
,\;k=1,2,\ldots ,N,  \nonumber \\
d_{jk} &=&\sqrt{\frac{2}{N+1}}\sin \left( \frac{kj\pi }{N+1}\right)
,\;j=1,2,\ldots ,N.  \label{eq:Huckel_sols}
\end{eqnarray}
If we choose $Q_{1}=\sqrt{\beta /\alpha }$ then it follows from equation (%
\ref{eq:Q_j^2}) that 
\begin{equation}
c_{jk}=\left( \frac{\beta }{\alpha }\right) ^{j/2}d_{jk}.
\label{eq:c_j->d_j_Huckel}
\end{equation}
Note that the eigenvalues and eigenvectors satisfy the general result
derived in section~\ref{sec:lattice} for the case $\omega _{j}=\omega $.

In a recent paper, Yuce\cite{Y21} studied a particular case of the model (%
\ref{eq:H}) with $\alpha _{j}=1$, $\beta _{j}=\gamma $, $\omega
_{j}=(-1)^{j}iV_{0}$ and $N$ even and stated that ``obtaining the spectrum
analytically is challenging''. The eigenvalue equation for this model is
exactly solvable when $\gamma =1$\cite{MM21}. Although $\omega _{j}$ is
complex we can nevertheless apply the approach developed in section~\ref
{sec:lattice} and obtain 
\begin{equation}
d_{j-1}+\left[ (-1)^{j}iv_{0}-y\right] d_{j}+d_{j+1}=0,\;v_{0}=\frac{V_{0}}{%
\sqrt{\gamma }},\;y=\frac{E}{\sqrt{\gamma }},  \label{eq:secular_complex}
\end{equation}
that is exactly solvable\cite{MM21}. We thus conclude that the energies are
given by 
\begin{equation}
E_{k}^{2}=4\gamma \cos ^{2}\left( \frac{k\pi }{N+1}\right)
-V_{0}^{2},\;k=1,2,\ldots ,\frac{N}{2},  \label{eq:E_k_sec_comp}
\end{equation}
which are real provided that 
\begin{equation}
\left| V_{0}\right| <2\sqrt{\gamma }\cos \left( \frac{N\pi }{2\left[
N+1\right] }\right) .  \label{eq:V_0_crit}
\end{equation}
This expression predicts that there are no real energies when $N\rightarrow
\infty $.

\section{Conclusions}

\label{sec:conclusions}

We have proved that if $\alpha _{j}$, $\beta _{j}$ and $\omega _{j}$ are
real and $\alpha _{j}\beta _{j}>0$, then the Hamiltonian (\ref{eq:H}) is
similar to a Hermitian operator and therefore its eigenvalues are real. The
reason for starting from the secular equation (\ref{eq:secular}) is that our
proof was motivated by an earlier argument proposed by Child et al\cite
{CDW00} and Amore and Fern\'{a}ndez\cite{AF20} for the truncation of
tridiagonal recurrence relations coming from the application of the
Frobenius method to quasi-solvable quantum-mechanical models. In this paper,
that mathematical strategy has proved useful for the construction of a
similarity transformation between a non-Hermitian and a Hermitian
Hamiltonian and has also enabled us to solve the eigenvalue equation for a
supposedly non-solvable model. It is worth noting that in the present case
of a finite-dimensional vector space the terms Hermitian operator,
self-adjoint operator and symmetric operator are equivalent.

Finally, we want to mention that the results developed in section~\ref
{sec:lattice} can be slightly generalized as shown in the appendix~\ref
{sec:Appendix}.

\appendix

\section{Generalization}

\label{sec:Appendix}

\renewcommand{\theequation}{A.\arabic{equation}} \setcounter{equation}{0}

In this Appendix we generalize the results developed in section~\ref
{sec:lattice}.

A linear operator $H$ can be expanded in a complete orthonormal basis set $%
\left\{ \left| j\right\rangle \right\} $ as 
\begin{equation}
H=\sum_{i}\sum_{j}H_{i,j}\left| i\right\rangle \left\langle j\right|
,\;H_{i,j}=\left\langle i\right| H\left| j\right\rangle .  \label{eq:H_gen}
\end{equation}
Assume that $\left| \psi \right\rangle $ is an eigenvector of $H$ with
eigenvalue $E$, $H\left| \psi \right\rangle =E\left| \psi \right\rangle $,
and carry out the transformation 
\begin{equation}
\left| \psi \right\rangle =Q\left| \varphi \right\rangle
,\;Q=\sum_{j}Q_{j}\left| j\right\rangle \left\langle j\right| ,\;Q_{j}\neq 0,
\label{eq:Q}
\end{equation}
so that 
\begin{equation}
\tilde{H}\left| \varphi \right\rangle =E\left| \varphi \right\rangle ,\;%
\tilde{H}=Q^{-1}HQ.  \label{eq:H_gen_transf}
\end{equation}
Therefore, the matrix elements of the transformed operator $\tilde{H}$ are 
\begin{equation}
\tilde{H}_{i,j}=Q_{i}^{-1}Q_{j}H_{i,j}.  \label{eq:H_(i,j)_gen_transf}
\end{equation}

If we restrict the analysis to a tri-diagonal Hamiltonian 
\begin{equation}
H=\sum_{j}\left( H_{j,j+1}\left| j\right\rangle \left\langle j+1\right|
+H_{j+1,j}\left| j+1\right\rangle \left\langle j\right| +H_{j,j}\left|
j\right\rangle \left\langle j\right| \right) ,  \label{eq:H_tri_diag}
\end{equation}
and require that $\tilde{H}_{j,j+1}=\tilde{H}_{j+1,j}^{*}$, then we derive
the relation 
\begin{equation}
\left| \frac{Q_{j+1}}{Q_{j}}\right| ^{2}=\frac{H_{j+1,j}^{*}}{H_{j,j+1}}%
=R_{j}>0.  \label{eq:relation_H_(i,j)}
\end{equation}
We conclude that any tridiagonal linear operator (\ref{eq:H_tri_diag}) that
satisfies $H_{i,i}=H_{i,i}^{*}$ and $H_{j+1,j}^{*}=R_{j}H_{j,j+1}$, $R_{j}>0$%
, is similar to an Hermitian operator of the form 
\begin{equation}
\tilde{H}=\sum_{j}\left( \sqrt{R_{j}}H_{j,j+1}\left| j\right\rangle
\left\langle j+1\right| +\frac{H_{j+1,j}}{\sqrt{R_{j}}}\left|
j+1\right\rangle \left\langle j\right| +H_{j,j}\left| j\right\rangle
\left\langle j\right| \right) ,  \label{eq:H_tri_diag_transf}
\end{equation}
where, for simplicity, we have chosen $Q_{j}=Q_{j}^{*}$. Therefore,
\begin{equation}
Q_{j}=\sqrt{R_{j-1}R_{j-2}\cdots R_{1}}Q_{1}.  \label{eq:Q_j}
\end{equation}

From the results above we can easily derive an additional condition for a
cyclic chain to be quasi-Hermitian. If we take into account that $%
H_{N,N+1}=H_{N,1}$ and $H_{N+1,N}=H_{1,N}$ we conclude that we have to add
the condition
\begin{equation}
H_{N,1}^{*}=R_{N-1}R_{N-2}\cdots R_{1}H_{1,N}.  \label{eq:cyclic_condition}
\end{equation}


\begin{thebibliography}{99}
\bibitem{BBJ03}  C. M. Bender, D. C. Brody, and H. F. Jones, Am. J. Phys. 
\textbf{71}, 1095 (2003).

\bibitem{B07}  C. M. Bender, Rep. Prog. Phys. \textbf{70}, 947 (2007).

\bibitem{Z07a}  M. Znojil, J. Phys. A \textbf{40}, 13131 (2007).

\bibitem{Z07b}  M. Znojil, J. Phys. A \textbf{40}, 4863 (2007).

\bibitem{Z07c}  M. Znojil, J. Phys. B \textbf{650}, 440 (2007).

\bibitem{Z08a}  M. Znojil, J. Phys. A \textbf{41}, 292002 (2008).

\bibitem{Z08b}  M. Znojil, Phys. Lett. D \textbf{78}, 025026 (2008).

\bibitem{Z09}  M. Znojil, Phys. Lett. D \textbf{80}, 045022 (2009).

\bibitem{Z10}  M. Znojil, Phys. Rev. A \textbf{82}, 052113 (2010).

\bibitem{Z11}  M. Znojil, J. Phys. A \textbf{44}, 075302 (2011).

\bibitem{P43}  W. Pauli, Rev. Mod. Phys. \textbf{15}, 175 (1943).

\bibitem{SGH92}  F. G. Scholtz, H. B. Geyer, and F. J. W. Hahne, Ann. Phys. 
\textbf{213}, 74 (1992).

\bibitem{W69}  J. P. Williams, Proc. Amer. Math. Soc. \textbf{20}, 121
(1969).

\bibitem{P68}  F. L. Pilar, Elementary Quantum Chemistry (McGraw-Hill, New
York, 1968).

\bibitem{CDW00}  M. S. Child, S-H. Dong, and X-G. Wang, J. Phys. A \textbf{33%
}, 5653 (2000).

\bibitem{AF20}  P. Amore and F. M. Fern\'{a}ndez, Phys. Scr. \textbf{95},
105201 (2020). arXiv:2007.03448 [quant-ph]

\bibitem{Y21}  C. Yuce, Phys. Lett. A \textbf{403}, 127384 (2021).

\bibitem{MM21}  R. Modak and B. P. Mandal, Phys. Rev. A \textbf{103}, 062416
(2021).
\end{thebibliography}
\end{document}